\documentclass[showpacs,preprintnumbers,amsmath,amssymb]{revtex4}
\usepackage{epsfig}
\usepackage{graphicx}

\newcommand{\itv}[1]{\mbox{\boldmath ${#1}$}}
\begin{document}
\draft

\title{Exact density matrix of the Gutzwiller wave function:
II. Minority spin component 
}

\author{Onuttom Narayan, Yoshio Kuramoto* and Mitsuhiro Arikawa**}
\affiliation{
Department of Physics, University of California, Santa Cruz, CA 95064\\
*Department of Physics, Tohoku University, Sendai, 980-8578, Japan\\
**Institute of Physics,  University of Tsukuba, Tsukuba, 305-8571, Japan
}

\date{\today}

\begin{abstract}
The density matrix, {\it i.e.} the Fourier transform of the momentum
distribution, 
is obtained analytically for all magnetization 
of the Gutzwiller wave function in one dimension with exclusion of double occupancy per site.
The present result complements the previous analytic derivation of the density matrix for the majority spin. 
The derivation makes use of a determinantal form of the squared wave function, and multiple integrals over particle coordinates are performed with the help of a diagrammatic representation.
In the thermodynamic limit,  the density matrix at distance $x$ is completely characterized by quantities $v_c x$ and $v_s x$, where $v_s$ and $v_c$ are
spin and charge velocities in the supersymmetric {\it t-J} model for which
the Gutzwiller wave function gives the exact ground state.
The present result then gives 
the exact density matrix of the {\it t-J} model for all densities and all magnetization at zero temperature.  Discontinuity, 
slope, and curvature singularities in the momentum distribution are identified.  
The momentum distribution obtained by numerical Fourier transform is in excellent agreement with existing result.

\end{abstract}


\maketitle
\section{Introduction}

The Gutzwiller wave function is the simplest many-particle wave function capable of capturing both itinerant and localized characters of strongly correlated
electrons \cite{gutzwiller}.  
For a long time, it has been considered as
a variational wave function for standard models such as the Hubbard
model and the {\it t-J} model \cite{brinkman,metzner}.  
It is now known that the Gutzwiller wave
function without double occupancy for each site constitutes the exact
ground state for the supersymmetric {\it t-J} model in one dimension,
for which both the exchange and transfer terms in the Hamiltonian decay
as inverse square of the distance \cite{kuramoto91}.
Given this context, it is highly meaningful to derive exact properties
for the Gutzwiller wave function without double occupation per site.
Various quantities have been studied exactly such as the spin
and charge correlation functions, and the momentum distribution
which is the Fourier transform of the density matrix.  Although the
correlation functions have been obtained in a simple closed
form \cite{forrester95, arikawa-saiga06}, most results for the momentum
distribution still involve a final  integration, or summation over
infinite series \cite{gebhard,kollar,arikawa04b}.

For a case with finite magnetization, Kollar and Vollhardt \cite{kollar}
extended the method of Ref.\cite{gebhard}, and obtained  the infinite
series expansion of the momentum distribution.  The expansion is valid
for general dimensions and general degree of restricting the double
occupation.  In one dimension they could sum up the series into a 
form involving a single integral over elliptic functions.  The resultant
mathematical form is very complicated, and it is difficult to obtain
any insight from the final formula in Ref.\cite{kollar}.

On the other hand, a completely different approach has been taken by
Arikawa et al \cite{arikawa04a,arikawa04b}.   Namely, the exact Green function
is integrated over the energy to give the momentum distribution.
The result has been obtained only for the singlet ground state, and
still involves integrals.  However, the structure of the integral makes
it clear how the elementary excitations of spin and charge determine
the momentum distribution.

In a previous paper \cite{ONYK}, hereafter referred to as I, we adopted another approach, and 
derived the exact density matrix of the Gutzwiller wave function in a closed form.
Our results in I, however, have
been restricted to the majority-spin component.
In this paper, we derive the minority spin component, completing the
analytic derivation for all magnetizations and all densities.   In the
singlet ground state without magnetization, the limit from the minority
spin agrees with that from the majority spin.  

Our strategy in I and this paper is to approach in the real space instead
of the momentum space.  In the real space,  all singularities in the momentum distribution appear in the asymptotic behavior of the density matrix. 
Hence we can avoid dividing the momentum space into different regions
separated by characteristic momenta. 
This paper gives the explicit form of the density matrix with use of Bessel functions.
The form obtained allows interpretation in terms of elementary excitations of
spin and charge.  
It is remarkable that different spin and charge velocities combine in such a way as to reproduce the asymptotic behavior expected from combination of the Fermi momenta of up and down spins.

\section{Gutzwiller wave function in Jastrow form}

We work with the particular representation of the Gutzwiller wave function 
with exclusion of double occupancy at each site.
We choose as the reference state the fully polarized state
$|F\rangle$ where each site is occupied by an up spin \cite{anderson89}.
Then a down spin at site $j$ is created by operating with
$S_j^-=S_j^x-iS_j^y$ on $|F\rangle$, and a hole by
$h_j^\dagger = c_{j\downarrow}S_j^-$.
Note that $h_j$ does not obey the ordinary anticommutation rule of fermions.
We can represent any state in terms of the wave function
$\Psi ( \{x^{\rm s}\}, \{x^{\rm h}\} )$ as \cite{anderson89}
\begin{equation}
| \Psi \rangle = \sum_{   \{x^{\rm s}\} ,  \{x^{\rm h}\}  } \Psi ( \{x^{\rm s}\}, \{x^{\rm h}\}  ) 
\prod_{i \in  \{x^{\rm s}\}  } S_{i}^- 
\prod_{j \in  \{x^{\rm h}\}  } h_{j}^\dagger | F \rangle,
\label{coordinate}
\end{equation}
where the set of coordinates $\{x^{\rm s}\}$ specifies the positions of
$M$ magnons, and $\{x^{\rm h}\}$ specifies those of $Q$ holes.
We call this scheme the magnon-hole representation.
The hard-core constraint is satisfied by the property
$
(S_j^-) ^2 = (h_j^\dagger) ^2 =S_j^-h_j^\dagger =0
$.
By definition
$\Psi( \{x^{\rm s}\}, \{x^{\rm h}\} )$ is symmetric against interchange
of down-spin coordinates, and antisymmetric against hole coordinates.
This is related to the commutation rule $S_i^- S_j^ -= S_j^-S_i^-$
and the anticommutation rule $h_i^\dagger h_j^\dagger = -h_j^\dagger
h_i^\dagger$ for $i\neq j$.
Furthermore, we have the relation $S_i^- h_j^\dagger
=h_j^\dagger S_i^- $.

The Gutzwiller wave function is represented in the magnon-hole representation by
\begin{equation}
\Psi_{\rm G}(\{x^{\rm s}\},\{x^{\rm h}\})  = 
\prod_{j=1}^{M+Q}\exp({\rm i}\pi x_j)
\prod_{i < j } D(x^{\rm s}_{i} -x^{\rm s}_{j})^2 
\prod_{l < m } D(x^{\rm h}_{l} -x^{\rm h}_{m})
\prod_{i, l } D(x^{\rm h}_{i} -x^{\rm s}_{l}),
\label{Gutzwiller-M}
\end{equation}
where $D(i-j) = (L/{\pi})\sin \left[ \pi\left(i-j\right)/L \right]$, 
$x_j$ denotes both hole $(j=1,\ldots, Q)$ and magnon
$(j=Q+1,\ldots,M+Q)$  coordinates.
This form is  valid for non-negative magnetization, {\it i.e.\/} $2M + Q \leq L.$
We have omitted a normalization factor here.
Here we have assumed the simplest case of 
$L$ and $M$ even, and $Q$ odd
\cite{note1},
for which the ground state is nondegenerate.
In the thermodynamic limit, the result does not depend on the choice of even or odd numbers of particles.
The momentum associated with $\Psi_{\rm G}(\{x^{\rm s}\},\{x^{\rm h}\})$ is $\pi$ because we have taken $M+Q$ odd.   However the reference state $|F\rangle$ itself has momentum $\pi$ since the occupied set of momentum includes one of the Brillouin zone boundary $\pi$, which does not cancel with the rest of occupied momentum.
Thus the Gutzwiller state $|\Psi_{\rm G}\rangle$ has zero momentum because of the cancellation of $\pi$ in $|F\rangle$ and $\Psi_{\rm G}(\{x^{\rm s}\},\{x^{\rm h}\})$.

It is convenient to introduce the complex coordinate
$z_j=\exp(2\pi i x_j/L)$ for both magnons and holes using the identity:
\begin{equation}
2i\sin\left[ \pi\left(x_i-x_j\right)/L \right]  =  (z_i-z_j)/\sqrt{z_i z_j}.
\end{equation}
Then the Gutzwiller wave function is given as a polynomial of these
complex coordinates, apart from the factor for the Galilean boost.
We work with the form
\begin{equation}
\Psi_{\rm G}(\{z\}) = \Psi_{\rm m+h} (\{z \})\Psi_{\rm m}(\{z\}),
\label{lowest}
\end{equation}
where the factors in the right-hand side are given by
\begin{equation}
\Psi_{\rm m+h} (\{z \})= i^{-(M+Q)(M+Q-1)/2}
\prod_{i=1}^{Q+M} z_i^{L/2-(Q+M-1)/2} \prod_{i<j}(z_i-z_j),
\label{hole}
\end{equation}
and
\begin{equation}
\Psi_{\rm m} (\{z\}) =i^{-M(M-1)/2}
\prod_{i=Q+1}^{Q+M} z_i^{-(M-1)/2}
\prod_{Q<i<j\leq Q+M} (z_i-z_j)
\label{Psi_m}
\end{equation}
where we have dropped factors of $L/(2\pi)$ from each $(z_i-z_j)$
as compared with Eq.(\ref{Gutzwiller-M}).

The Gutzwiller wave function given by Eq.(\ref{Gutzwiller-M})
turns out to give
the exact ground state of the {\it t-J} model with a special condition
for the transfer and the exchange interaction \cite{kuramoto91}.  The {\it t-J} model
in general is given by
\begin{eqnarray}
{\mathcal H}_{tJ}
&=&
\sum_{i<j}
{\cal P}
\left[-t_{ij}\sum_{\sigma=\uparrow,\downarrow}
\left( c^{\dagger}_{i\sigma}c_{j\sigma} + h.c. \right)
+J_{ij}
\left(\itv S_{i}\cdot\itv S_{j}
-\frac{1}{4} n_{i}n_{j}\right)
\right]
{\cal P},   \label{tj-hamiltonian}
\end{eqnarray}
where $c_{i\sigma}$ is the annihilation operator of an electron with spin
$\sigma$ at site $i$, $n_{i}$ is the number operator, and ${\cal P}$
is the projection operator to exclude double occupancy at each site.
We assume that  the parameters satisfy the following condition
\begin{equation}
t_{ij}=J_{ij}/2= t D (i-j)^{-2}
\label{1/r^2}
\end{equation}
This model is called the supersymmetric {\it t-J} model with
inverse-square interaction \cite{kuramoto91}.  Here the lattice constant
is taken as the unit of length.  Hence the length $L$ of the system
gives also the number of lattice sites.

\section{Density matrix in terms of the wave function}

We work with the
electron density matrix $\rho_\sigma(x)$ for 
spin up $(\sigma=\uparrow)$ and 
spin down $(\sigma=\downarrow)$ electrons, which
is the Fourier transform of the
momentum distribution function $n_\sigma (k)$.  
Thus we are interested in evaluating 
\begin{equation}
\rho_\sigma (x) = \langle c_\sigma^\dagger(x+x_j) c_\sigma(x_j)\rangle
\label{sigmaprop}
\end{equation} 
in the ground state. Here we have written 
$c_\sigma (x_j) $ for $c_{j\sigma}$.
The left hand side of the equations is independent
of $x_j$ because of translational invariance. 
In I, we have derived $\rho_\uparrow(x)$ analytically for a magnetization 
$m=n_\uparrow- n_\downarrow \ge 0$.   The other part $\rho_\downarrow(x)$ requires more elaborate calculation, which is presented in this paper. 

For the down spin part of the density matrix, we have to evaluate
the expectation value in the ground state, with $M+1$ magnons and $Q-1$ holes, of
\begin{eqnarray}
\rho_\downarrow (x) 
= \langle  b_x^\dagger  h_x h_0^\dagger b_0  \rangle
= -\langle  b_x^\dagger  h_0^\dagger h_x b_0  \rangle +\delta_{x,0}{{M+1}\over L}
\equiv -G_\downarrow (x) 
 +\delta_{x,0}{{M+1}\over L}
\label{Expect}
\end{eqnarray}
where the hole operators behave as fermionic.  
We will refer to $G_\downarrow(x)$
as the propagator.
Ignoring the Kronecker's $\delta$
in the last expression, the expectation value forces one of the
holes in the ket wavefunction to be at $x$ and one of the magnons
at the zeroth site, while in the bra wavefunction this is reversed.
All other holes and magnons have to be matched in the bra and ket
wavefunctions. 

Using Eq.(\ref{lowest}) with Eqs.(\ref{hole}) and
(\ref{Psi_m}), we use the normalization
\begin{equation}
B(M,Q) = \int\frac{dx_1}{L}\ldots\frac{dx_{M+Q}}{L}
\Psi_{{\rm m}+{\rm h}}^2(z_1\ldots z_{M+Q})\Psi_{\rm m}^2(z_{Q+1}\ldots z_{M+Q}). 
\label{bmq}
\end{equation}
(Since $\Psi_{ {\rm m}+{\rm h}}$ and $\Psi_{\rm h}$ are products of real factors, complex
conjugation is not necessary.)
Therefore, evaluating the expectation value in Eq.(\ref{Expect}) in the
ground state with $Q-1$ holes and $M+1$ magnons, we have
\begin{eqnarray}
\langle b_x^\dagger h_0^\dagger h_x b_0\rangle &=&
\frac{1}{B(M+1,Q-1)} \frac{Q-1}{L}\frac{M+1}{L}
\frac{(1 - z_x)^2}{z_x}
\int\frac{dx_2}{L}\ldots\frac{dx_{Q-1}}{L}
   \frac{dx_{Q+1}}{L}\ldots\frac{dx_{Q+M}}{L}\nonumber\\
&\times &\Bigg(\prod_{i=2}^{Q-1}\frac{(1 - z_i)(z_x - z_i)}{-z_i\sqrt{z_x}}
\prod_{i=Q+1}^{Q+M} \frac{(1 - z_i)(z_x - z_i)}{-z_i\sqrt{z_x}}\Bigg)^2
\prod_{i=Q+1}^{Q+M}\frac{(z_x - z_i)(1 - z_i)}{-z_i\sqrt{z_x}}\nonumber\\
&\times &\Psi_{{\rm m}+{\rm h}}^2(z_2\ldots z_{Q-1}, z_{Q+1}\ldots z_{M+Q}) 
\Psi_{\rm m}^2(z_{Q+1}\ldots z_{M+Q})
\label{exp_coord}
\end{eqnarray}
where $z_x=\exp(2\pi ix/L)$, and
the factor $\Psi_{\rm m+h}^2 \Psi_{\rm m}^2$ represents a system
with $Q-2$ holes and $M$ magnons. 
Eq.(\ref{exp_coord})
can be written in terms of the wavefunction for $Q$ holes and $M$ magnons,
with the first two holes at the zeroth and $x$ sites, as
\begin{eqnarray}
G_\downarrow(x) &=&
-\frac{1}{B(M+1,Q-1)} \frac{Q-1}{L}\frac{M+1}{L}
\int\frac{dx_2}{L}\ldots\frac{dx_{Q-1}}{L}
   \frac{dx_{Q+1}}{L}\ldots\frac{dx_{Q+M}}{L}\nonumber\\
&\times &\prod_{i=Q+1}^{Q+M}\frac{(z_x - z_i)(z_i - 1)}{z_i\sqrt{z_x}}
\Psi_{{\rm m}+{\rm h}}^2(1, z_x, z_2\ldots z_{Q-1}, z_{Q+1}\ldots z_{M+Q}) 
\Psi_{\rm m}^2(z_{Q+1}\ldots z_{M+Q})
\label{exp_rewrite}
\end{eqnarray}
which further simplifies to
\begin{equation}
G_\downarrow(x) = 
-\frac{B(M,Q)}{B(M+1,Q-1)} \frac{Q-1}{L}\frac{M+1}{L}
\Bigg\langle (L\delta_{x_Q, 0}) (L\delta_{x_1,x})
\prod_{i=Q+1}^{Q+M}\frac{(z_x - z_i)(z_i - 1)}{z_i\sqrt{z_x}}\Bigg\rangle.
\label{expectation}
\end{equation}
In this last expression, the expectation value is evaluated in the
ground state with $M$ magnons and $Q$ holes, with hole coordinates at
$z_1\ldots z_Q$ and magnon coordinates at $z_{Q+1}\ldots z_{Q+M}.$ The
factors of $L$ associated with the Kronecker's $\delta$ makes them $O(1)$
in the thermodynamic limit.

In Eq.(\ref{expectation}), there is a factor for each magnon coordinate 
which can be expanded as
\begin{equation}
2\cos\phi - \exp[i\phi]/z_i - z_i\exp[-i\phi],
\label{poly_factor}
\end{equation}
after defining
$z_x = \exp[2i\phi]$,
or $\phi = \pi x/L$.
The product of each of these factors can, inside the expectation value
of Eq.(\ref{expectation}), be expressed as a sum of diagrams, similar
to I.   This is pursued further in the next section.

\section{Evaluation of propagator in diagrams}

\subsection{Basics of the diagram technique}

We first review how the normalization $B(M,Q)$ is evaluated. This discussion is  
almost identical to that in Ref.\cite{ONYK}, but is reproduced here because it
is essential to understanding the diagrammatic expansion. 
We make extensive use of a determinant representation of  $|\Psi_{\rm
G}(\{z\})|^2$ following ref.\cite{forrester95}.   We introduce a
notation \begin{equation} \det_V(z_1\ldots ,z_Q) \equiv \det(z_1^p\ldots
,z_Q^p)_{p=-(Q-1)/2\ldots (Q-1)/2}, \label{Vandermonde} \end{equation}
where the suffix $V$ means the Vandermonde determinant.  In the matrix
for the determinant, each row has an entry $(z_1^p\ldots z_Q^p)$ with
$p=-(Q-1)/2$ in the first row, and $p=(Q-1)/2$ in the $Q$-th (last) row.
Furthermore, the confluent alternant is introduced by
\begin{equation}
\det_A(z_1\ldots z_{M+Q}) \equiv \det(z_1^p\ldots z_Q^p,z_{Q+1}^p, p z_{Q+1}^p
    \ldots z_{M+Q}^p, p z_{M+Q}^p)_{p = - (2M + Q - 1)/2\ldots (2M + Q - 1)/2}.
\label{alternant}
\end{equation}
It can be shown that $\det_A(z_1\ldots z_{M+Q})$ has fourth-order
zeros $(z_i-z_j)^4$ if both $z_i$ and $z_j$ are magnon coordinates,
second-order zeros $(z_i-z_j)^2$ if one of them is a hole coordinate,
and first-order zeros $(z_i-z_j)$ if both $z_i$ and $z_j$ are hole
coordinates.  Since $\Psi_{\rm G}$ in Eq.(\ref{Gutzwiller-M}) and in
Eq.(\ref{lowest}) is real, we obtain up to a real positive factor
\begin{equation}
|\Psi_{\rm G} (\{z\})|^2 = \Psi_{\rm G}(\{z\})^2 =
(-1)^{P(M,Q)}
\det_V(z_1\ldots z_Q) \det_A(z_1\ldots z_{M+Q}).
\label{square}
\end{equation}
Here $P(M,Q) = M(M-1)/2 + (M+Q)(M+Q-1)/2$ comes from the factors of $(-i)$
in Eqs.(\ref{hole}) and (\ref{Psi_m}). One can verify that there
is no additional factor of $(-1)$ from the determinants;
 taking the
diagonal term from both determinants, which has a coefficient of $+1,$
one obtains the lowest power
possible for $z_1,$ followed by the next lowest power for $z_2,$ and
so on. 
On the other hand, 
in the polynomials in Eq.(\ref{hole}) and
(\ref{Psi_m}), since
each $(z_i - z_j)$ has $i<j,$ 
this term is obtained by taking $-z_j$
from each such factor. 
Since each such factor in $\Psi_{\rm G}$ is
repeated in $\Psi^2_{\rm G},$ we have an overall coefficient of $+1.$
The coefficient of this is therefore also $+1,$
and there is no factor of $(-1)$ in going from the polynomials to the
determinants.
Since $(-1)^n = 1$ for any even
$n,$ $P(M,Q)$ is equivalent to $[Q^2 + Q(2M - 1)]/2 \equiv MQ + Q(Q-1)/2.$

Eq.(\ref{bmq}) can now be written as
\begin{equation}
B(M,Q) = (-1)^{P(M,Q)}\int \frac{dx_1}{L}\ldots\frac{dx_{M+Q}}{L}
\det_V(z_1,z_2\ldots z_Q) \det_A(z_1\ldots z_{M+Q}).
\label{normal}
\end{equation}
The first determinant is a sum of terms 
of the form $z_1^{p_1}\ldots z_Q^{p_Q},$ and the second determinant is a sum
of terms of the form $z_1^{q_1}\ldots z_Q^{q_Q} 
(p_{Q+1} - q_{Q+1}) z_{Q+1}^{p_{Q+1} + q_{Q+1}}\ldots
(p_{Q+M} - q_{Q+M}) z_{Q+M}^{p_{Q+M} + q_{Q+M}},$ where $\{p_1\ldots p_Q\}$
is a permutation of $-(Q-1)/2\ldots (Q-1)/2$, 
 and $\{q_1\ldots q_{Q+M},p_{Q+1}
\ldots p_{Q+M}\}$ is a permutation of $-(Q+2M-1)/2\ldots (Q+2M-1)/2.$ 
We adopt the convention that $p_i > q_i$ for the magnons. 
Here $p_i$'s and $q_i$'s are the momenta of the holes and magnons. 
Integrating the hole coordinates, $z_1\ldots z_Q,$ 
we see that $p_i + q_i = 0,$  
i.e. the $p_i$'s and $q_i$'s are equal and opposite, both
covering the range $[-(Q-1)/2, (Q-1)/2].$ Now integrating the magnon
coordinates, $z_{Q+1}\ldots z_{M+Q},$ we see that $p_i + q_i=0$ for 
the magnons too. Since the hole coordinates cover $[-(Q-1)/2,(Q-1)/2],$
the magnon $p_i$'s range from $(Q+1)/2$ to $(Q+2M-1)/2$,
and the minus sign counterpart.

\begin{figure}
\centerline{\epsfxsize=4in\epsfbox{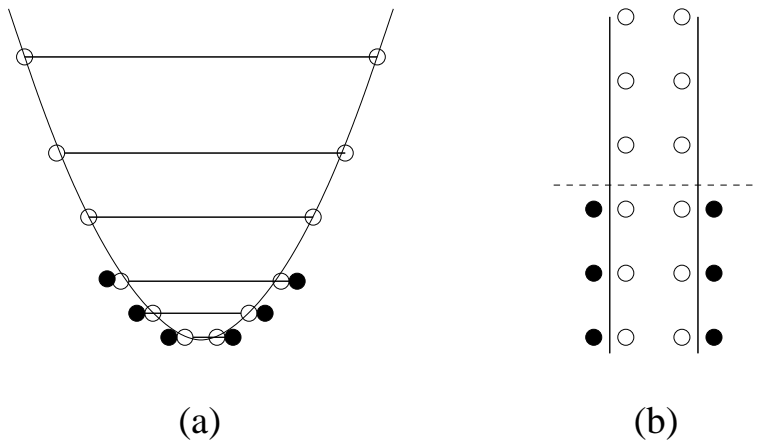}}
\caption{(a) 
Diagram for evaluating the product of Vandermonde and alternant
determinants.  The white circles denote the exponents in the
alternant determinant, integer spaced from $-(Q+2M-1)/2$ to $(Q+2M-1)/2.$
The horizontal positions of the circles give the exponents, which are
equally spaced horizontally. 
The black circles denote the exponents in the Vandermonde determinant, and are slightly separated from the white circles for clarity. 
A line connecting a black circle to a white one represents a hole
coordinate, and one connecting a white circle to a white one represents
a magnon coordinate.  
For clarity, some of the hole lines have not been shown: every black circle should
be connected to the white circle on the opposite side at the same level.
(b)  Schematic of the same diagram, where a vertical projection has been
taken and the points are equally spaced on the vertical axis. The actual
horizontal separation between two points can be calculated by counting how many levels down one side and up the other separate them. The dashed line is the magnon hole boundary.} 
\label{fig1} 
\end{figure}

Diagrammatically, $B(M,Q)$ can be represented as in Fig.\ref{fig1}, where 
$p_i + q_i = 0$ forces all the lines to be horizontal. The holes are represented
by the lines connecting black circles (from the Vandermonde determinant) to
white circles (from the alternant), while the magnons are represented by lines
connecting white circles to white circles. Every magnon line contributes a
factor of $p_i - q_i,$ i.e. its horizontal extent, while every hole line
contributes a factor of unity.  Various permutation symmetry factors have
to be included to calculate $B(M,Q)$. The final result is 
\begin{equation}
B(M,Q) = M!Q! (Q+1)(Q+3)\ldots (Q+2M-1).
\label{BMQ}
\end{equation}
As a result of this, Eq.(\ref{expectation}) can be written as 
\begin{equation}
G_\downarrow(x) = -A(M,Q) 
\langle (L\delta_{x_Q, 0}) (L\delta_{x_1,x})
\prod_{i=Q+1}^{Q+M}\left( 
2\cos\phi - e^{i\phi}/z_i - z_i e^{-i\phi} \right)\rangle
\label{expect}
\end{equation}
with
\begin{equation}
A(M,Q) = \frac{M+1}{L}\frac{Q-1}{L} \frac{B(M,Q)}{B(M+1,Q-1)} =
\frac{Q(Q-1)}{L^2}\frac{(Q+1)(Q+3)\ldots(Q+2M-1)}{Q(Q+2)\ldots(Q+2M)}.
\label{amq}
\end{equation}

Expanding the product of factors in Eq.(\ref{expect}), for each magnon
coordinate we have a factor of $2\cos\phi,$  $-\exp[i\phi]/z_i$ or its
inverse. Therefore in the diagrams contributing to the expectation value
in Eq.(\ref{expect}), the total momentum for each magnon is 0 or $\pm
1,$ while the total momentum for each hole is still zero. 
Here we have loosely referred these integers to ``momentum", although physical momentum 
should be multiplied by $2\pi/L$.
In addition,
two of the holes --- the first and the $Q$'th ones --- have no constraint
on their momentum, because of the Kronecker's $\delta$ in coordinate space.
In Fig.~\ref{fig1}, each magnon line connects two white circles, but
can now be horizontal, ascending one step (going from left to right) or
descending one step. Each hole line connects a black circle to a white
circle, and is horizontal. However, for two of the holes the black circle
can connect to a white circle at any level. We refer to these hereafter
as ``jokers", because their momenta can assume any value. For clarity,
in subsequent figures the black circles are replaced by hatched circles
for the jokers. Note that if the joker at $x=x$ connects to a line that 
ascends $n$ levels, it implies that the diagram has an overall factor
of $z_x^n = \exp[2in\phi].$ On the other hand, for the joker at $x=0$ 
an ascending line yields a factor of unity. This is in contrast to the 
magnon lines, where an ascending line is associated with a factor of
$\exp[i \phi],$ not $\exp[2i\phi].$ 
As discussed at the beginning of this section, each magnon line also 
contributes a factor equal to its total momentum.

As a warmup exercise, we evaluate $\langle (L\delta_{x_Q, 1})
(L\delta_{x_1,x}) (2\cos\phi)^M\rangle$ in Eq.(\ref{expect}). All
the magnon lines are horizontal, and have the same contribution as in
$B(M,Q).$ Therefore either each joker connects to the white circle at its
level, or the jokers exchange partners. In the first case, the momenta
for both the jokers are
zero, as in $B(M,Q).$ Instead of the sum over
$x_0$ and $x_Q,$ we have a factor of $L$ with each Kronecker's $\delta$,
which has the same effect. In the second case, there is an extra factor
of $-[\exp(2i\phi)]^{q_Q - q_1},$ where the $(-1)$ factor comes from
the permutation of terms within the determinants. The expectation value
is therefore
\begin{equation} 
\frac{(2\cos\phi)^M}{Q(Q-1)} 
\sum_{q_1\neq q_Q}
\left[ 
1 -\exp(2i\phi)^{q_Q - q_1}
 \right]
\end{equation}
which sums to 
\begin{equation} 
(2\cos\phi)^M \Bigg[1 - 
\frac{1}{Q(Q-1)}\Bigg\{\frac{\sin^2 Q\phi}{\sin^2\phi} - Q\Bigg\}\Bigg].  
\label{warmup}
\end{equation}

We are now in a position to evaluate the diagrams in Eq.(\ref{expect}). 
We classify the diagrams into four categories, in increasing order of 
complexity,
and evaluate each of them. As in Ref.~\cite{ONYK}, we 
refer to the boundary between magnon and hole momenta in $B(M,Q)$ as 
the magnon-hole boundary, shown by a dashed line in Fig.~\ref{fig1},
even though in a general diagram there can be magnon lines that go
below this boundary and hole lines that go above this (for the holes 
that are the two jokers).

\subsection{Type I diagrams}
In these diagrams, the jokers pair with the white circles at their own
level, or exchange partners. Therefore neither joker connects to a white
circle above the magnon hole boundary. All the other holes have horizontal
connections. This is a generalization of the diagrams just evaluated for
$\langle (L\delta_{x_1,x})(L\delta_{x_Q,0})\rangle,$ except that 
now the magnon lines need not be horizontal. Since all the hole lines 
stay below the magnon hole boundary, the magnon lines have to stay above 
it. As explained in Ref.\cite{ONYK}, each magnon line is either horizontal
or exchanges partners with an adjacent magnon line to form a `dimer'. 
As a result, the factor of $(2\cos\phi)^M$ changes to $F(M,Q),$ 
introduced in I.
$F(M,Q)$ describes the contribution of magnon pairs and dimers, and is defined
 by the recursion relation
\begin{equation}
F(M, Q) = 2\cos\phi F(M - 1, Q + 2) - [1 + 1/((Q + 2)^2 - 1)] F(M - 2, Q + 4),
\end{equation}
with the boundary conditions $F(0, Q + 2 M) = 1$ and 
$F(1, Q + 2 M - 2) = 2 \cos\phi.$ 
From Eq.(\ref{warmup}),
the contribution to the propagator from type I diagrams is then
\begin{equation}
G_I(x) = -A(M,Q) F(M,Q) \Bigg[1 - 
\frac{1}{Q(Q-1)}\Bigg\{\frac{\sin^2 Q\phi}{\sin^2\phi} - Q\Bigg\}\Bigg].  
\label{g_I}
\end{equation}

\begin{figure}
\centerline{\epsfxsize=3in\epsfbox{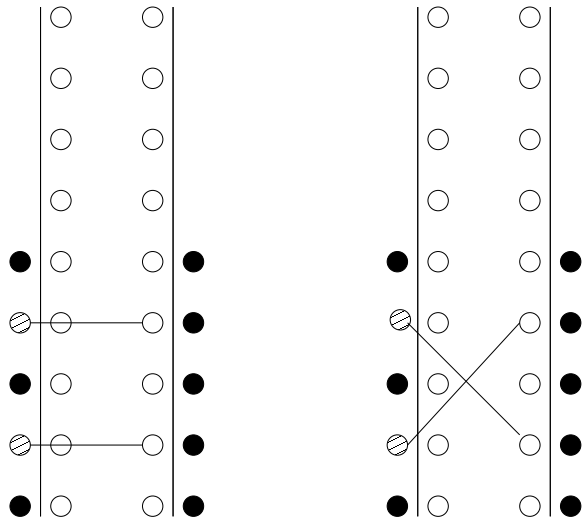}}
\caption{Type-I diagrams. The jokers have shaded black circles.
The two diagrams shown in the figure are a complementary pair. 
Both jokers are partnered with white circles below the magnon hold
boundary. All magnon lines are horizontal or dimerized.}
\label{fig11} 
\end{figure}
The Type I diagrams can be divided into two complementary classes: for
every diagram where the lines from the two jokers are horizontal, there
is a a corresponding diagram where their partners are exchanged. This
generalizes to the subsequent, more complicated, classes of diagrams:
diagrams occur in complementary pairs, where the two diagrams in a pair
differ in that
the partners of the jokers are exchanged. The exchange of partners causes a factor of
$-\exp[2 i (q_Q - q_1)\phi],$ where $q_Q - q_1$ depends on the diagram. 
If $\phi=0,$ the
diagrams cancel each other, and all contributions to $G_\downarrow(x)$ are
zero. This is reasonable, since $G_\downarrow(x)$ involves destroying a hole and
a magnon in the ground state, and the same site cannot have both a hole
and a magnon.

\subsection{Type II diagrams}

\begin{figure}
\centerline{\epsfxsize=3in\epsfbox{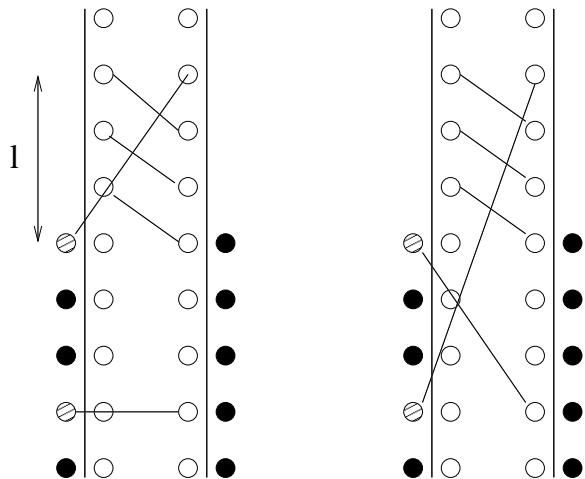}}
\caption{Type-II diagrams. 
The jokers have shaded black circles.
The two diagrams shown in the figure are a complementary pair. 
One joker is partnered with a white circle above the line, $l$ levels
above the magnon hole boundary. All magnon lines that are not shown
are above this, and are horizontal or dimerized.}
\label{fig2} 
\end{figure}

In these diagrams, one joker partners with a white circle above the
magnon hole boundary. The white circle that it would have paired with
now has to be partnered. Recalling that, except for the jokers, holes
must be linked with horizontal lines, this can happen in one of two
ways. Either the other joker pairs with the white circle at its level, in
which case the first joker must be just below the magnon hole boundary,
and the magnon line just above the boundary moves down one level. Alternatively,
we have the same situation with the partners of the jokers exchanged. The
two possibilities together are the complementary pair of diagrams just discussed,
and are shown in Fig.~\ref{fig2}. Since magnon lines can only have a
total momentum of $\pm 1$ or 0, if the joker links to a circle $l$ levels
above the magnon hole boundary, this affects the $l$ magnons immediately
above the boundary, as shown in the figure.  Beyond the $l$'th level,
the magnon lines behave as before: they are horizontal or dimerized.

The contribution from these diagrams to the propagator is
\begin{eqnarray}
G_{II}(x) &&= -A(M,Q)\sum_{l=1}^M {1\over Q} F(M-l, Q + 2l)
\frac{Q}{Q+1}\frac{Q+2}{Q+3}\ldots\frac{Q+2l-2}{Q+2l-1}\nonumber\\
&\times &\Bigg[\exp(2il\phi)\exp(-il\phi)
   \Big\{1 - \frac{1}{Q-1}\sum_{p=1}^{Q-1}\exp\big(-2i(l+p)\phi\big)\Big\}
   \nonumber\\
&+& \exp(il\phi)\exp(-2il\phi)
   \Big\{1 - \frac{1}{Q-1}\sum_{p=1}^{Q-1}\exp\big(2i(l+p)\phi\big)\Big\}
    + \phi\leftrightarrow -\phi\Bigg].\nonumber\\
\label{g_II}
\end{eqnarray}
The factors inside the summation are explained as follows. Let us
first consider the case when the joker at $x=x$ is just below
the left edge of the magnon hole boundary, shown in the first panel
in Fig.~\ref{fig2}, and links to a white circle $l$ levels above the
magnon hole boundary.  The factor of $1/Q$ is because this joker could
have occupied any of the $Q$ black circles below the magnon hole boundary,
and is now restricted to one position. The factor of $F(M-l, Q+2l)$ is
because beyond the $l$'th level above the magnon hole boundary, the magnon
lines are horizontal or dimerize as in Type I diagrams. The contribution
from these magnons is as if the magnon hole boundary was moved up by $l$
levels. The factor of $Q/(Q+1)\ldots (Q+2l - 2)/(Q+2l-1)$ is because
each magnon line contributes a factor equal to its total momentum. 
The lowest $l$ magnon lines, which are no longer horizontal, have each
been reduced in total momentum by unity.  The factor of $\exp(2il\phi)$
is because the joker at $x=x$ has a total momentum of $l,$ resulting in
a factor of $(z_x)^l = \exp(2il\phi).$ The factor of $\exp(-il\phi)$
comes because $l$ magnon lines each have total momentum $-1,$ from the
Vandermonde and alternant determinants, so they must each have taken a
factor of $-\exp(-i\phi)$ from Eq.(\ref{poly_factor}), and there is an
additional $(-1)^l$ permutation factor from the determinants because the
line from the joker goes over $l$ magnon lines it was previously under.

The next case is the group of diagrams complementary to the ones
just considered, shown in the second panel in Fig.~\ref{fig2}. This
group gives rise to the second term inside the first $\{\}$ brackets in
Eq.(\ref{g_II}). If the second joker is $p$ levels away from the first
one, $p$ is summed from 1 to $Q-1,$ with a weighting factor of $1/(Q-1)$
for each. The total momentum of the first joker changes by $-(l+p)$
relative to the first diagram from the exchange of partners,
resulting in a factor of $-\exp(-2i(l+p)\phi).$ 

We also have to consider diagrams similar to these two classes, but 
in which the first joker is just below the right end of the magnon
hole boundary.  Now the total momentum of the joker is $-l,$ and the
$l$ lowest magnon lines are ascending rather than descending. For the
complementary diagrams, the total momentum
of the first joker changes by $(l+p).$ 

The remaining diagrams are similar to the four classes considered so
far, except that the first joker is the one at $x=0$ rather than the
one at $x=x.$ It is possible to verify for each group separately
that this changes $\phi$ to $-\phi.$ All these contributions add up
to Eq.(\ref{g_II}). It is possible to verify that the 
$\phi\leftrightarrow -\phi$ is equivalent to an overall factor of 2.

The sums over $p$ in Eq.(\ref{g_II}) can be performed, yielding
\begin{eqnarray}
G_{II}(x) &&= -2 A(M,Q)\sum_{l=1}^M F(M-l, Q + 2l)
\frac{(Q+2)(Q+4)\ldots (Q + 2 l - 2)}{(Q + 1)(Q+3)\ldots(Q + 2 l - 1)}\nonumber\\
&\times & \Bigg\{\exp(il\phi)[1 - \frac{\exp(iQ\phi)}{Q-1}\frac{\sin(Q-1)\phi}{\sin\phi}]
   + \phi\leftrightarrow - \phi\Bigg\}.
\label{G_II_int}
\end{eqnarray}

\subsection{Type III diagrams}

\begin{figure}
\centerline{\epsfxsize=2in\epsfbox{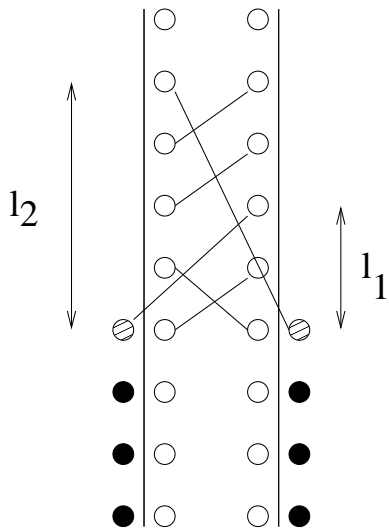}}
\caption{Type-III diagrams. 
Both jokers are partnered with white circles above the line, 
$l_1$ and $l_2$ levels above the magnon hole boundary. 
All magnon lines that are not shown are below or above the 
$[l_1, l_2]$ block, and are horizontal or dimerized.}
\label{fig3} 
\end{figure}

Moving on in order of increasing complexity, we consider diagrams in which
both jokers link to white circles above the magnon hole boundary. The
white circles they would have paired with need partners. For Type III
diagrams, we assume that these partners come from above the magnon hole
boundary. This requires that both jokers should be immediately below the 
boundary. 
(For
Type IV diagrams, we will assume that these two white circles pair with
each other.)  By an extension of the argument for Type II diagrams,
both the jokers have to be immediately below the magnon hole boundary,
on either side of it. We assume that the jokers to the left and right
pair with white circles $l_1$ and $l_2$ levels above the magnon
hole boundary respectively. As seen in Fig.~\ref{fig3}, both white
circles immediately above the magnon hole boundary connect to circles
below the boundary, so that $l_1$ and $l_2$ have to be no less than
2. There is a block of ascending (for $l_2 > l_1$) or descending
(for $l_1 > l_2$) magnon lines from $l_{\min}$ to $l_{\max},$ where
$l_{\min}$ and $l_{\max}$ are the lesser and greater of $l_1$ and $l_2$
respectively. All other magnon lines are either horizontal or form dimers.

The total contribution to the propagator from Type III diagrams is
\begin{eqnarray}
G_{III}(x) &=& A(M,Q)\sum_{l_1 = 2}^M\sum_{l_2 = 2}^M
\frac{1}{Q(Q-1)} F(l_{\min} - 2, Q + 2) F(M - l_{\max}, Q + 2
      l_{\max})\nonumber\\
&\times& \frac{Q^2}{Q+1}
\frac{(Q+2l_{\min})(Q + 2 l_{\min} + 2)\ldots(Q+2l_{\max} - 2)}
{(Q+2l_{\min} - 1)(Q + 2 l_{\min} + 1)\ldots(Q+2 l_{\max} - 1)}\nonumber\\
&\times& \exp[i (l_2 - l_1)\phi]
\Bigg\{\exp(2 i l_1\phi)\Big[1 - \exp(-2 i(Q - 1 + l_1 + l_2)\phi)\Big]
      + \exp(-2il_2\phi)\Big[1 - \exp(2 i (Q - 1 + l_1 + l_2)\phi)\Big]
\Bigg\}.\nonumber\\
\label{g_III'}
\end{eqnarray}
In this equation, there is a factor of $(-1)$ from the permutation of 
the determinants, which cancels the $(-1)$ in front.
The $1/Q(Q-1)$ is because the location of the two jokers
is fixed as being immediately below the magnon hole boundary. 
The factors of $F$ come from the dimerized magnon lines below
$l_{\min}$ and above $l_{\max}.$ The factors of $(Q+{\rm const})$ are from
the magnon lines: instead of one line of length $Q+1$ immediately above
the magnon hole boundary, there are two lines of length $Q$ that cross
the boundary, and instead of $l_{\max} - l_{\min} + 1$ horizontal magnon
lines from the $l_{\min}$ to the $l_{\max}$ levels, there are $l_{\max}
- l_{\min}$ ascending or descending lines. The ascending or descending
lines also contribute $l_2 - l_1$ factors of 
$-e^{i\phi}$ or $l_1 -
l_2$ factors of $-\exp[-i\phi]$ respectively; the $(-1)$'s are absorbed
in the permutation factor at the beginning. The $\exp(2i l_1\phi)$
comes from diagrams shown in Fig.~\ref{fig3},
with the $x=x$ joker at the left end of the magnon hole boundary.
The complementary diagrams have an 
additional $-\exp(-2i(Q-1+l_1 + l_2)\phi).$ The remaining
terms are from the same diagrams, but with the $x=x$ joker at the
right end of the magnon hole boundary. Eq.(\ref{g_III'}) simplifies to
\begin{eqnarray}
G_{III}(x) &=& A(M,Q)\sum_{l_1 = 2}^M\sum_{l_2 = 2}^M
\frac{Q}{Q^2 -1} F(l_{\min} - 2, Q + 2) F(M - l_{\max}, Q + 2
      l_{\max})\nonumber\\
&\times &
\frac{(Q+2l_{\min})(Q + 2 l_{\min} + 2)\ldots(Q+2l_{\max} - 2)}
{(Q+2l_{\min} - 1)(Q + 2 l_{\min} + 1)\ldots(Q+2 l_{\max} - 1)}\nonumber\\
&\times &
\Bigg\{2\cos((l_1+l_2)\phi) - 2\cos((l_1 + l_2 + 2 Q - 2)\phi)
\Bigg\}.\nonumber\\
\label{g_III}
\end{eqnarray}

\subsection{Type IV diagrams}

\begin{figure}
\centerline{\epsfxsize=3in\epsfbox{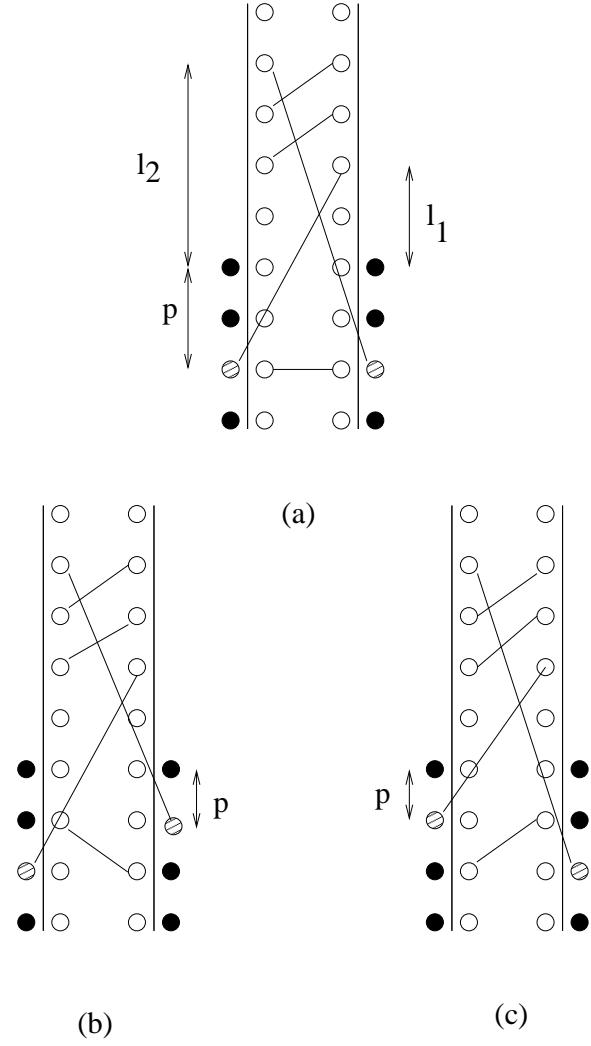}}
\caption{Type-IV diagrams. 
Both jokers are partnered with white circles above the line, 
$l_1$ and $l_2$ levels above the magnon hole boundary. 
The white circles that they leave unpaired connect with each
other, forming a magnon line below the magnon hole boundary.
This line must be horizontal, or ascending or descending by one 
level. All three cases are shown in the figure.}
\label{fig4} 
\end{figure}

Finally, we come to diagrams where both jokers connect to white circles
above the line, and the white circles they would have partnered with
(i.e. at their level) pair with each other to form a magnon line. This
is shown in Fig.~\ref{fig4}. The two jokers now no longer have to be
immediately below the magnon hole boundary, but can be at any arbitrary
depth below it. However, in order for their white circles to pair and form
a magnon line, both jokers must be at the same level, or differ in level
by unity. The three possible cases are shown in Fig.~\ref{fig4}. The
contributions to the propagator from the first case is
\begin{eqnarray}
G_{IVa}(x) &=& -A(M,Q)\sum_{p=0}^{Q/2 - 1} \sum_{l_1 = 1}^M \sum_{l_2 = 1}^M
(2\cos\phi)\frac{1}{Q(Q-1)} F(l_{\min} - 1, Q) 
   F(M - l_{\max}, Q + 2 l_{\max})\nonumber\\
&\times &(Q-1-2p) \frac{(Q+2 l_{\min})\ldots(Q+2 l_{\max} - 2)}
{(Q + 2 l_{\min} - 1)\ldots(Q + 2 l_{\max} - 1)}\nonumber\\
&\times &
\exp[i(l_2 - l_1)\phi]
\Bigg\{\exp(2i(l_1 + p)\phi)\Big[1 - \exp(-2 i (l_1 + l_2 + Q - 1)\phi)\Big]
   \nonumber\\
   &+& \exp(-2i(l_2 + p)\phi)\Big[1 - \exp[ 2 i (l_1 + l_2 + Q - 1)\phi)\Big]
      \Bigg\}.\nonumber\\
\label{g_IVa}
\end{eqnarray}
This is similar to $G_{III}(x),$ so we only discuss the
differences. $p$ is the number of levels below the magnon hole boundary
that the jokers are placed.  The factor of $2\cos\phi$ comes from
the magnon line below the magnon hole boundary, as does the factor of
$Q-1-2p.$ The four terms within the brackets come from diagrams in the
first panel of Fig.~\ref{fig4} with the $x=x$ joker on the left hand
side, the complementary group of diagrams, and the same with the $x=x$
joker on the right hand side.

The contributions from the second and third cases are 
\begin{eqnarray}
G_{IVb}(x) &=& A(M,Q)\sum_{p=0}^{Q/2 - 2} \sum_{l_1 = 1}^M \sum_{l_2 = 1}^M
\frac{1}{Q(Q-1)} F(l_{\min} - 1, Q) 
   F(M - l_{\max}, Q + 2 l_{\max})\nonumber\\
&\times &(Q-2-2p) \frac{(Q+2 l_{\min})\ldots(Q+2 l_{\max} - 2)}
{(Q + 2 l_{\min} - 1)\ldots(Q + 2 l_{\max} - 1)}\nonumber\\
&\times &
\exp[i(l_2 - l_1 - 1)\phi]
\Bigg\{\exp(2i(l_1 + p + 1)\phi)\Big[1 - \exp(-2 i (l_1 + l_2 + Q - 1)\phi)\Big]
   \nonumber\\
   &+& \exp(-2i(l_2 + p)\phi)\Big[1 - \exp[ 2 i (l_1 + l_2 + Q - 1)\phi)\Big]
      \Bigg\}\nonumber\\
\label{g_IVb}
\end{eqnarray}
and 
\begin{eqnarray}
G_{IVc}(x) &=& A(M,Q)\sum_{p=0}^{Q/2 - 2} \sum_{l_1 = 1}^M \sum_{l_2 = 1}^M
\frac{1}{Q(Q-1)} F(l_{\min} - 1, Q) 
   F(M - l_{\max}, Q + 2 l_{\max})\nonumber\\
&\times &(Q-2-2p) \frac{(Q+2 l_{\min})\ldots(Q+2 l_{\max} - 2)}
{(Q + 2 l_{\min} - 1)\ldots(Q + 2 l_{\max} - 1)}\nonumber\\
&\times &
\exp[i(l_2 - l_1 + 1)\phi]
\Bigg\{\exp(2i(l_1 + p)\phi)\Big[1 - \exp(-2 i (l_1 + l_2 + Q - 1)\phi)\Big]
   \nonumber\\
   &+& \exp(-2i(l_2 + p + 1)\phi)\Big[1 - \exp[ 2 i (l_1 + l_2 + Q - 1)\phi)\Big].
      \Bigg\}\nonumber\\
\label{g_IVc}
\end{eqnarray}
Here $p$ is the distance of the joker that is closer to the magnon hole
boundary from the boundary. The differences between these and $G_{IVa}$
can be explained.  The sum over $p$ stops at $Q/2 - 2,$ because the lower
joker is $p+1$ levels below the magnon hole boundary. There is an overall
minus sign relative to $G_{IVa}$
which is a permutation factor. $2\cos\phi$ is replaced by
$\exp(-i\phi)$ and $\exp(i\phi)$ respectively, because the magnon line
below the magnon hole boundary is descending or ascending instead of
being horizontal. Finally, inside the brackets, we have $\exp(2i(l_1 +
p + 1)\phi)$ in $G_{IVb}$ because the joker on the left is $l_1 + p +
1$ levels below its partner, i.e. has total momentum $l_1 + p + 1.$
(The $\exp(-2i(l_2 + p + 1)\phi)$ in $G_{IVc}$ is similar.)

It is tedious but straightforward to carry out the sum over $p$ in
Eqs.(\ref{g_IVa}), (\ref{g_IVb}) and (\ref{g_IVc}). All three can be
added up, and yield
\begin{eqnarray}
G_{IV}(x) &=& -A(M,Q)\sum_{l_1 = 1}^M \sum_{l_2 = 1}^M
\frac{1}{Q} F(l_{\min} - 1, Q) 
   F(M - l_{\max}, Q + 2 l_{\max})\nonumber\\
&\times &\frac{(Q+2 l_{\min})\ldots(Q+2 l_{\max} - 2)}
{(Q + 2 l_{\min} - 1)\ldots(Q + 2 l_{\max} - 1)}\nonumber\\
&\times &\Bigg\{2\cos((l_1 + l_2 - 1)\phi) - 2 \cos((l_1 + l_2 + 2 Q - 1)\phi)\Bigg\}.
\label{g_IV}
\end{eqnarray}

\section{Thermodynamic limit}
\subsection{Synthesis of all contributions to $G_\downarrow$ }

The expressions we have obtained for the various parts of $G_\downarrow(x)$
simplify in the thermodynamic limit $L\rightarrow\infty$.
This limit makes $\phi \rightarrow 0$, but
$Q\phi$ and $M\phi$ remain finite, and are given by 
\begin{align}
Q\phi = v_c x,\quad
(Q+2M)\phi = v_s x.
\end{align}
Here we have fixed
$Q/L= 1-n$ and $M/L=(n-m)/2$, and used
the velocities 
$v_c=\pi (1-n)$ for charge and 
$v_s=\pi (1-m)$ for spin.
Elementary excitations with these velocities appear in the supersymmetric {\it t-J} model where both exchange and transfer decay as the inverse square of the distance \cite{kuramoto91}. 
Also, we obtain from Eq.(\ref{amq}) in the thermodynamic
limit,
\begin{equation}
A(M,Q) \rightarrow 
\frac{Q}{L^2}\sqrt{\frac{Q}{Q+2M}} =
\frac{1-n}{L}\sqrt{\frac{v_c}{v_s}} 
\label{AMQ}
\end{equation}
and 
\begin{equation}
\frac{(Q+2)\ldots(Q+2l-2)}{(Q+1)(Q+3)\ldots(Q+2l-1)} \rightarrow
\frac{1}{\sqrt{(Q+1)(Q+2l-1)}}
\label{Ql}
\end{equation}
if $l/L$ is finite. Finally, from Ref.~\cite{ONYK}, in the thermodynamic limit
\begin{align}
F(M,Q) &= \frac{\pi}{4}\sqrt{Q(Q+2M)}
\Bigg[Y_0(\frac{Q+2M}{2}\phi)J_0(\frac{Q}{2}\phi) - \left( J\leftrightarrow Y\right)\Bigg]\notag\\
&=
\frac\pi 4 L^2\sqrt{v_cv_s}\left[ 
Y_0(\frac12 v_s x)J_0(\frac12 v_c x)-
J_0(\frac12 v_s x)Y_0(\frac12 v_c x)
 \right]
\label{FMQ}
\end{align}
where $J_\nu$ and $Y_\nu$ are Bessel functions of $\nu$-th order.
Using Eqs.(\ref{AMQ}), (\ref{FMQ}), (\ref{Ql}) and (\ref{g_I}), we obtain
in the thermodynamic limit
\begin{equation}
G_I(x) \rightarrow 
 -\frac{v_c^2}{4 \pi} \left[ Y_0(\frac{v_s}{2} x) J_0 (\frac{v_c}{2} x)
-J_0(\frac{v_s}{2} x) Y_0 (\frac{v_c}{2} x)\right]
\left( 1- \frac{\sin^2 v_c x}{v_c^2 x^2}\right).
\label{G_I}
\end{equation}

Next we proceed to $G_{II}(x)$.
In Eq.(\ref{G_II_int}), replacing the sum over $l$ with an integral 
with $y = (l+Q/2)\phi$, 
and applying
Eqs.(\ref{AMQ}), (\ref{FMQ}), and (\ref{Ql}), we obtain in the thermodynamic limit
\begin{align}
G_{II}(x)\rightarrow -\frac{v_c}{\pi x}\Re \int_{v_cx/2}^{v_sx/2} dy
\left[ 
Y_0(\frac{v_s x}{2}) J_0( y) e^{iy}- \left( J\leftrightarrow Y\right) \right]
\left[ 
e^{-iv_c x/2}-e^{iv_c x/2}
\frac{\sin v_c x}{v_c x} \right],
\label{G_II}
\end{align}
where $\Re$ takes the real part.
Here we have used the form for $F(M-l,Q+2l)$ that is
valid if $(Q+2l)/L$ and $(M-l)/L$ are both finite in the thermodynamic
limit.
This is correct over essentially the entire range of $l$ as
$L\rightarrow\infty.$
We see that the expressions in 
Eqs.(\ref{G_I}) and (\ref{G_II}) are finite in the thermodynamic limit. 
As given in Eq.(\ref{diff}),  we
use the indefinite integral result
\begin{equation}
\int dy e^{iy}  Z_0(y) = \left[ 
Z_0(y) + i Z_0^\prime(y) \right]ye^{iy}
\label{besselint}
\end{equation}
where $Z_0 = J_0$ or $Y_0$ is a Bessel function of zeroth order.
Then integration of Eq.(\ref{G_II}) results in
\begin{align}
G_{II}(x) 
& =  -\frac{2v_c}{\pi^2 x}\left[ \sin \frac{v_s-v_c}{2} x - \sin \frac{v_s+v_c}{2} x \frac{\sin v_c x}{v_c x}\right] 
\nonumber \\
& 
 + \frac{v_c^2}{2\pi} \left[ Y_0(\frac{v_s x}{2}) J_0(\frac{v_c x}{2})-
J_0(\frac{v_s x}{2}) Y_0(\frac{v_c x}{2})\right] \left( 1- \frac{\sin v_c x \cos v_c x}{v_c x}\right) 
\nonumber \\
& 
 - \frac{v_c \sin^2 (v_c x)}{2\pi x} \left[ Y_0(\frac{v_s x}{2}) J_1(\frac{v_c x}{2})-
J_0(\frac{v_s x}{2}) Y_1(\frac{v_c x}{2})\right].  
\label{G_ii}
\end{align}
where we have used the relation 
$J'_0(y)=-J_1(y)$ and  $Y'_0(y)=-Y_1(y)$
given by Eq.(\ref{diff}), and the simple Wronskian given by Eq.(\ref{Wronskian}).

We continue on to evaluation of 
$G_{III}(x)$ and $G_{IV}(x)$, which are more complicated. 
From Eqs.(\ref{g_III}) and (\ref{g_IV}), the thermodynamic limits of 
both $G_{III}(x)$ and $G_{IV}(x)$ are infinite:
$A(M,Q)$ is $O(1/L),$ the two $F$ functions are both $O(L)$ if all their
arguments are $O(L),$ as they generically are, and the polynomials in $Q$
in the numerator and denominator have an overall $O(1/L^2)$ behavior. The
double sum, over $l_1$ and $l_2,$ makes $G_{III}(x)$ and $G_{IV}(x)$
of $O(L).$
However, $G_{III}$ and $G_{IV}$ almost exactly cancel each other,
and their difference is finite in the thermodynamic limit. Comparing
Eqs.(\ref{g_III}) and (\ref{g_IV}), we see that $G_{III}(x)$ and
$G_{IV}(x)$ are equal and opposite, except for four differences:
(i) the sums over $l_1$ and $l_2$ have different ranges, (ii) the polynomials in
$Q$ in the numerator and denominator are different, (iii) the arguments of the
first $F$ function are different, and (iv) the arguments of the cosines in the
brackets are different. 
Of these, (i) and (ii) cause negligible difference in the thermodynamic limit,
but (iii) and (iv) are significant.
We consider the effects of all four separately.

First, if $l_1 = 1,$ a term that exists in $G_{IV}(x)$ but not
$G_{III}(x),$ there is no $O(L)$ factor from the sum over $l_1,$
and $F(0,Q) = 1$ instead of being $O(L).$ Therefore the contribution
of this term to $G_{IV}(x)$ is $O(1/L)$ and can be neglected. The
same is true for $l_2 = 1.$ 
Second, there is a factor of $Q/(Q^2-1)$ in $G_{III}$ instead of
$1/Q$ in $G_{IV}.$ This is equal to $(1/Q)[1 + 1/Q^2 + O(1/Q^4)].$
The correction from this is 
$
G_{III}(x)/Q^2,$ which is $O(1/L)$
in the thermodynamic limit and can be neglected.

To account for the difference (iii), we note that the contribution for the sum of $G_{III}$ and $G_{IV}$ contains the term
\begin{align}
F(M - l_1, Q + 2 l_1)\left[ 
F(l_2 - 2, Q + 2)-F(l_2  -1, Q) \right],
\end{align}
where we have taken $l_{\min}=l_2$ and $l_{\max}=l_1$ without loss of generality.
In the thermodynamic limit, the difference $\Delta F$ of the two $F$'s is given by
\begin{align}
\Delta F= -\frac\pi 4\sqrt{\frac{2 y_2}{v_c}} \left\{ 
\left[ Y_0 (\frac 12 v_c x) J_0(y_2) -(J\leftrightarrow Y) \right]
+v_c x \left[ Y_0 (\frac 12 v_c x) J_0'(y_2) -(J\leftrightarrow Y) \right] \right\},
\end{align}
where $y_2=(Q/2+l_2)\phi$.  Then the whole contribution from (iii) is given by
\begin{align}
G_a (x)& \equiv \frac{\pi^2}{4 L^2}\sum_{l_1 = 1}^M\sum_{l_2 = 1}^{l_1}
[Y_0(\frac 12 v_s x) J_0(y_1)
- \left( J\leftrightarrow Y\right)]\left\{ 
Y_0(y_2)\left[ 
J_0(\frac 12 v_c x) + v_c x J_0^\prime(\frac 12 v_c x) \right]
- \left( J\leftrightarrow Y\right)
 \right\}
\nonumber\\
&\times
 \Re e^{i(y_1+y_2)}\left( 
e^{-iv_c x}-e^{iv_c x}
\right) 
\label{G_III_IVb}
\end{align}
where $y_1 = (Q+l_1/2)\phi$.
Replacing the sums with integrals 
over $y_1$ and $y_2$,
we obtain
\begin{align}
G_a(x) &= 
\frac{\sin (v_cx)}{2 x^2} \Im \int_{v_cx/2}^{v_sx/2} dy_1
\int_{v_cx/2}^{y_1} dy_2\left[ 
Y_0(\frac 12 v_s x) J_0(y_1)e^{iy_1}
- \left( J\leftrightarrow Y\right) \right]
\nonumber\\
&\times
\left\{ 
\left[ 
J_0(\frac 12 v_c x) + v_c x J_0^\prime(\frac 12 v_c x) \right]Y_0(y_2)e^{iy_2}
- \left( J\leftrightarrow Y\right)
 \right\}.
\label{Ga}
\end{align}
where $\Im$ takes the imaginary part.
Finally, the effect (iv) of the change in the argument of the cosines is to 
differentiate them, since $\phi\rightarrow 0,$ yielding 
\begin{align}
G_b(x) \equiv -\frac{\pi^2 v_c x}{4 L^2}\sum_{l_1=1}^M\sum_{l_2=1}^{l_1}
[Y_0(\frac 12 v_s x) J_0(y_1) - \left( J\leftrightarrow Y\right)]
[Y_0(y_2) J_0(\frac 12 v_c x) - \left( J\leftrightarrow Y\right)]
\Im 
e^{i(y_1+y_2)}\left( e^{-iv_c x}+e^{iv_c x} \right).
\label{G_III_IVc}
\end{align}
As in the case of $G_a$, replacing the sums with integrals 
over $y_1$ and $y_2$, we obtain
\begin{align}
G_b(x) =
\frac{v_c \cos (v_c x)}{2x}\Im
\int_{v_c x/2}^{v_sx/2} d y_1
\int_{v_c x/2}^{y_1} d y_2
\left[ 
Y_0 (\frac{v_s}{2} x) J_0 (y_1)e^{iy_1}
-(J \leftrightarrow Y) \right]
\left[ 
Y_0(\frac{v_c}{2} x)J_0(y_2) e^{iy_2}
-(J \leftrightarrow Y) \right] .
\label{Gb}
\end{align}
Then we obtain the thermodynamic limit of 
$
G_{III}(x)  +G_{IV}(x)  = 
G_{a}(x)  +G_{b}(x)  
$.

In Eqs.(\ref{Ga}) and (\ref{Gb}), 
integration over $y_1$ and $y_2$  can be performed analytically by the use of 
Eq.(\ref{besselint}).
The integral over $y_2$ has the upper limit $y_1$ and the lower limit $v_cx/2$.
Surprisingly, the contribution from the lower limit 
cancels out $G_{II}(x)$ exactly.
The final integration over $y_1$ with terms coming from the upper limit $y_2=y_1$ can be carried out by using the following formula, which are derived in Appendix: 
\begin{align}
I_{JJ}&\equiv
\int ^{v_s x/2}_{v_c x/2} d y \ y J_0(y) \left[ J_0(y) \sin 2y -J_1(y) \cos 2y  \right] 
 = \frac{1}{2} \Im \left[  (J_0(y) - i J_1(y) )^2 y^2e^{i2y}
\right]\Big|^{v_s x/2}_{v_c x/2} \label{I_JJ}
\\
I_{YY}&\equiv
\int ^{v_s x/2}_{v_c x/2} d y \ y Y_0(y) \left[ Y_0(y) \sin 2y -Y_1(y) \cos 2y  \right] 
 = \frac{1}{2} \Im \left[ (Y_0(y) - i Y_1(y) )^2 y^2e^{i2y}
\right]\Big|^{v_s x/2}_{v_c x/2} \\
I_{JY}&\equiv
\int ^{v_s x/2}_{v_c x/2} d y \ y J_0(y) \left[ Y_0(y) \sin 2y -Y_1(y) \cos 2y  \right]
\nonumber\\
&=\frac{1}{2} \Im \left[(J_0(y) - i J_1(y) )(Y_0(y) - i Y_1(y) )  y^2e^{i2y}
\right]\Big|^{v_s x/2}_{v_c x/2} + \frac{\sin v_s x - \sin v_c x}{2\pi} \\
I_{YJ}&\equiv 
\int ^{v_s x/2}_{v_c x/2} d y \ y Y_0(y) \left[ J_0(y) \sin 2y -J_1(y) \cos 2y  \right]
\nonumber\\
& =\frac{1}{2} \Im \left[ (Y_0(y) - i Y_1(y) )(J_0(y) - i J_1(y) ) y^2 e^{i2y}
\right]\Big|^{v_s x/2}_{v_c x/2} - \frac{\sin v_s x - \sin v_c x}{2\pi}.
\label{I_YJ}
\end{align}  
Collecting the terms, 
we find tremendous cancellation.   Furthermore, many combinations 
take the form of the Wronskian Eq.(\ref{Wronskian}), and 
can be simplified.
The propagator 
$G_\downarrow(x) = G_I(x)+G_{II}(x)+G_a(x)+G_b(x)$
is now given a closed form: 
\begin{align}
G_\downarrow(x) &=
-\frac{1}{4\pi x^2} 
\left[v_c x \cos(v_c x) - \sin(v_c x)\right] \left[v_s x \cos(v_s x) - \sin(v_s x)\right]
\left[J_0(\frac 12 v_s x) Y_0(\frac 12 v_c x) - Y_0(\frac 12 v_s x) J_0(\frac 12 v_c x)\right]
\notag\\
&- \frac{1}{4\pi } v_c \sin(v_c x) v_s \sin(v_s x)
\left[J_1(\frac 12 v_s x) Y_1(\frac 12 v_c x) - Y_1(\frac 12 v_s x) J_1(\frac 12 v_c x)\right]
\notag\\
&+\frac{1}{4 \pi x} v_c \sin(v_c x) \left[\sin(v_s x) - v_s x \cos(v_s x)\right]
\left[J_0(\frac 12 v_s x) Y_1(\frac 12 v_c x) - Y_0(\frac 12 v_s x) J_1(\frac 12 v_c x)\right]
\notag\\
&-\frac{1}{4 \pi x} v_s \sin(v_s x) \left[\sin(v_c x) - v_c x \cos(v_c x)\right]
\left[J_0(\frac 12 v_c x) Y_1(\frac 12 v_s x) - Y_0(\frac 12 v_c x) J_1(\frac 12 v_s x)\right].
\label{main}
\end{align}
This is the main result of the present paper.

\begin{figure}[t]
\includegraphics[width=0.59\linewidth,clip]{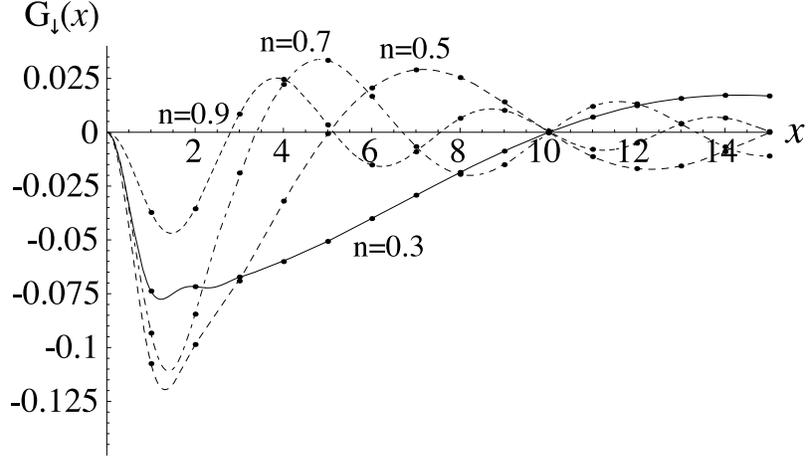}
\caption{
Distance dependence of $G_\downarrow(x)= -\rho_\downarrow (x)+\delta_{x,0}(n-m)/2$
with $n=0.9, 0.7, 0.5, 0.3$, and $m=0.1$ for all cases.
Only integer values of $x$ are physically relevant.
For comparison, the data obtained by inverse Fourier transform of Kollar-Vollhardt results are also shown by points.
}
\label{Gnm}
\end{figure}

Figure \ref{Gnm} shows 
$G_\downarrow(x)$ for a few cases of $n$ with $m$ fixed as $0.1$.
In order to test the results,  we have also computed the inverse Fourier transform of
Ref.\cite{kollar} to the real space, as shown by dots.  
The agreement between the present calculation and Ref.\cite{kollar} is excellent.
Although the curves in Fig.\ref{Gnm} are shown as continuous, 
and all pass through the origin because 
$G_\downarrow(x\rightarrow 0) = 0$ from Eq.(\ref{main}),
the coordinate $x$ is only physical for integer $x.$ 
This point will next be discussed in detail.

\subsection{Comparison between $G_\uparrow$ and $G_\downarrow$}

The up-spin density matrix is related to the propagator 
$G_\uparrow (x)=\langle h_0^\dagger h_x   \rangle 
$ introduced in I as 
\begin{align}
\rho_\uparrow (x) \equiv
\langle c_\uparrow^\dagger(x+x_j) c_\uparrow (x_j)\rangle
=\delta_{x,0}\left[ 1-\frac 12 (n-m) \right] -G_\uparrow (x).
\label{rho-up}
\end{align}
From the definition we obtain $G_\uparrow (x=0) = 1-n$, and 
$\rho_\uparrow (x=0) = (n+m)/2$.
On the other hand, using Eq.(\ref{Expect}) we obtain 
$\rho_\downarrow(x=0) = (n-m)/2$ and $G_\downarrow(x=0) = 0.$ We note that
$G_\sigma(x=0) = G_\sigma(x\rightarrow 0)$ for both up and down spin 
despite the fact that only integer
values of $x$ are physical. 

In the singlet case $m=0$, the up- and down-spin density matrices 
should coincide. This implies that 
$
G_\downarrow(x; m = 0) = G_\uparrow(x; m = 0)$
for $x\neq 0$, but  
\begin{align}
G_\downarrow(x=0; m = 0) \neq G_\uparrow(x=0; m = 0), 
\end{align}
because of a difference in the Kronecker's $\delta$ terms. 
Let us see how these relations appear in our result.
At $v_s = \pi$, which happens for $m=0$, 
$G_\downarrow (x;m=0)$ with integer $x\ (\neq 0)$ simplifies to
\begin{eqnarray}
G^{(r)}_\downarrow(x) & = & \frac{v_c (-1)^{x}}{4}
\left[ 
\left\{ 
\cos (v_c x) - \frac{\sin(v_c x) }{v_c x}   \right\}
\left\{
Y_0 \left( \frac{v_s x}{2} \right) J_0 \left( \frac{v_c x}{2} \right)
-
J_0 \left( \frac{v_s x}{2} \right) Y_0 \left( \frac{v_c x}{2} \right)
\right\} \right. \nonumber \\
&  & + \left. \sin (v_c x)
\left\{
Y_0 \left( \frac{v_s x}{2} \right) J_1 \left( \frac{v_c x}{2} \right)
-
J_0 \left( \frac{v_s x}{2} \right) Y_1 \left( \frac{v_c x}{2} \right)
\right\} \right]
\label{singlet-limit}
\end{eqnarray}
where the superscript in $G^{(r)}_\downarrow$ signifies that this reduced
form is only valid for $m=0$ and integer $x\neq 0.$
This indeed agrees with the result
in I for $G_\uparrow$ with $m=0$.
Thus the present calculation and the previous one consistently cover the whole range of magnetization.

\begin{figure}
\includegraphics[width=0.59\linewidth,clip]{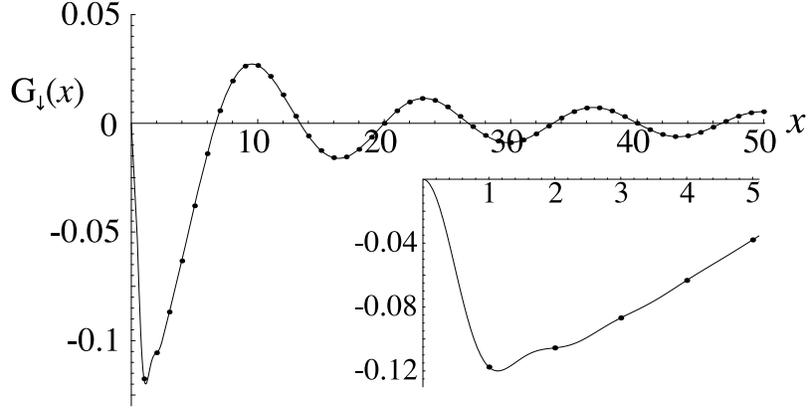}
\caption{
Distance dependence of $G_\downarrow(x)$
in the singlet limit with $n=0.3, \ m=0$ 
(solid line and dots).
The inset shows the expanded view for small $x$.
}
\label{n03m0}
\end{figure}

\begin{figure}
\includegraphics[width=0.59\linewidth,clip]{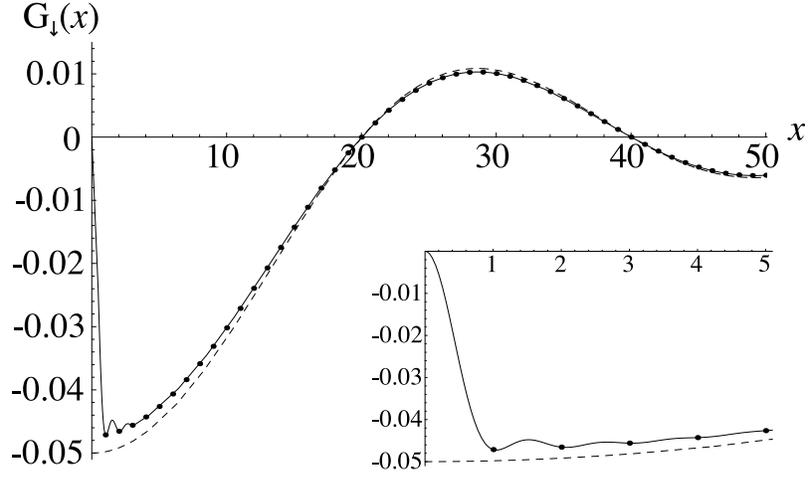}
\caption{
Distance dependence of $G_\downarrow(x)$
in the singlet limit with $n=0.1, \ m=0$ 
(solid line and dots).   
The shape is reasonably well fit
by $-\sin(\pi nx/2)/(\pi x)$ (dashed line), which is the case for non-interacting
electrons.
The inset shows the expanded view for small $x$.
}
\label{n01m0}
\end{figure}

Figure \ref{n03m0} shows the result for 
$G_\downarrow (x)$ given by Eq.(\ref{main})
with $n=0.3$ and $m=0$.  
The result for non-integer values of $x$ is included to clarify the analytic property, especially for small $x$. 
The curve for $x>1$ tends to the value $-\rho_\downarrow (0)=
-n/2=-0.15$, but finally goes to zero in the limit $x\rightarrow 0,$   
as discussed for Fig.~\ref{Gnm}.
The period $\Delta x$ of damped oscillation is given by $\Delta x\sim 4/n\sim 13$.  
One might naively expect from the trigonometric function in Eq.(\ref{singlet-limit}) that
\begin{align}
\Delta x \mathop{=}^? \frac{2\pi}{v_c}= \frac{4}{1-n},
\end{align}
which is not the case.
This situation becomes clearer in Fig. \ref{n01m0}, which shows the result for very dilute 
density $n=0.1$.  Since the electron correlation is not important except for small $x$, the 
result for $G_\downarrow (x)$ is well fit by the noninteracting result:
$G(x)= -\sin(\pi nx/2)/(\pi x)$.  
However, the hard-core constraint causes deviation from the noninteracting behavior near 
$x=0$.
The period of damped oscillation is determined by the Fermi momentum $k_F
= \pi n/2$, which is the same as the Fermi velocity in our unit.  
The Fermi velocity determines the principal oscillation for a general case of $n$ such as $n=0.3$.

\begin{figure}[t]
\includegraphics[width=0.59\linewidth,clip]{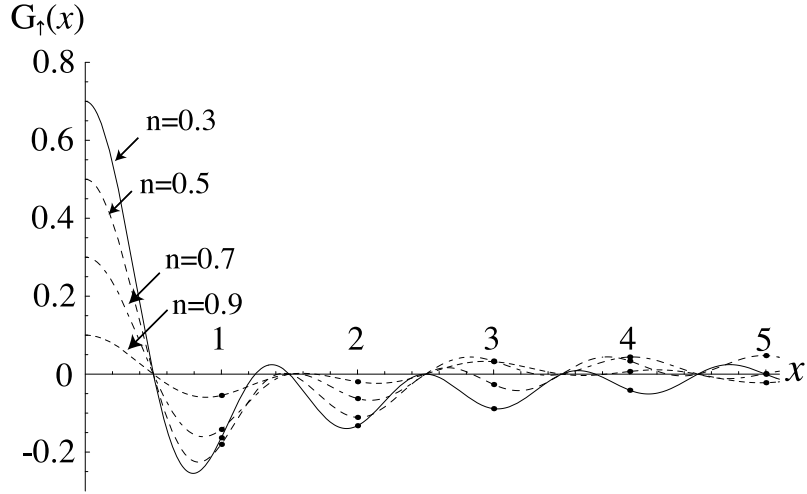}
\caption{
Distance dependence of 
$G_\uparrow(x)$
with $n=0.9, 0.7, 0.5, 0.3$, and $m=0.1$ for all cases.
Only integer values of $x$ are physically relevant,   
and the lines are shown only for easy tracking of $G_\uparrow(x)$ at integers.
}
\label{upspin}
\end{figure}

For comparison between $G_\uparrow$ and $G_\downarrow$,
Fig. \ref{upspin} shows 
$G_\uparrow (x)$ given by
\begin{align}
G_\uparrow(x)  = & \frac{v_c \cos(\pi x)}{4}
\left[ 
\left( \cos (v_c x) - \frac{\sin(v_c x) }{v_c x}  \right)
\left\{
Y_0 \left( \frac{v_s x}{2} \right) J_0 \left( \frac{v_c x}{2} \right)
-
J_0 \left( \frac{v_s x}{2} \right) Y_0 \left( \frac{v_c x}{2} \right)
\right\} \right. \nonumber \\
&   + \left. \sin (v_c x)
\left\{
Y_0 \left( \frac{v_s x}{2} \right) J_1 \left( \frac{v_c x}{2} \right)
-
J_0 \left( \frac{v_s x}{2} \right) Y_1 \left( \frac{v_c x}{2} \right)
\right\} \right],
\end{align}
where $\cos(\pi x)$ is used instead of $(-1)^x$ used in I, in order to
include non-integer values of $x$. 
The main difference from $G_\downarrow (x)$ is the finite limiting value
of $1-n$ as $x\rightarrow 0$. There are also
oscillations between neighboring integer values of $x,$ unlike for
$G_\downarrow(x),$ but these are not physical.
We note that, after replacing $(-1)^x$ with $\cos(\pi x),$ 
$G^{(r)}_\downarrow(x)$ is the same as $G_\uparrow(x),$ and therefore
has the same 
non-zero $x\rightarrow 0$ limit and oscillations, emphasizing 
the limited equivalence between $G^{(r)}_\downarrow(x)$ and $G_\uparrow(x).$

\subsection{Asymptotic behavior}

The long-distance behavior of the propagator shows the singularity of the momentum distribution, as discussed in I.  
Using the known asymptotic form of the Bessel functions,
and using the Fermi velocities
$k_{F\uparrow} = \pi(n+m)/2$ and $k_{F\downarrow} = \pi (n-m)/2$,
we have derived the leading behavior of $G_\downarrow (x)$ for large $x$ as
\begin{align}
G_\downarrow (x) & \sim
\frac{-1}{\pi x} \sqrt{(1-n)(1-m)} 
\left[ 
\sin k_{F\downarrow} x 
- \frac{n_\downarrow \cos k_{F\downarrow} x }{2\pi (1-n)(1-m) x} 
- \frac{(n-m)^2}{32\pi^2 (1-m)^2 (1-n)^2 x^2 } \sin k_{F\downarrow} x 
\right. \nonumber \\
&  \left. 
+ \frac{\sin [(2k_{F\uparrow}+k_{F\downarrow}) x]}{4\pi^2(1-n)^2 x^2}
- \frac{\sin [(2k_{F\uparrow}-k_{F\downarrow}) x]}{4\pi^2(1-m)^2 x^2}
\right] 
\label{down-asymp}
\end{align}
up to $O(1/x^3)$.
For reference, our previous result in I 
for the up spin is 
quoted \cite{note2} to $O(1/x^3)$ as
\begin{equation}
G_\uparrow (x) 
\sim {-1\over{\pi x}}\sqrt{{1-n}\over{1-m}}
\Bigg[\sin k_{F\uparrow}x 
- 
{ n_\downarrow \cos k_{F\uparrow} x \over{2\pi (1-n)(1-m)x}}  
- { {1 + n_\downarrow^2/[2(1-n)^2]}\over{4\pi^2(1-m)^2 x^2}}  
\sin k_{F\uparrow} x 
+ 
{ {\sin (k_{F\uparrow}+2k_{F\downarrow})x}\over{4\pi^2 (1-n)^2x^2}}\Bigg].
\label{k_Fx}
\end{equation}
In the case of singlet $m=0$,  the present result Eq.(\ref{down-asymp}) agrees with Eq.(\ref{k_Fx}).

\subsection{Singularities in momentum distribution}

The momentum distribution $n_\downarrow (k)$ is obtained from the Fourier transform of $G_\downarrow (x)$ as
\begin{align}
n_\downarrow(k) = \frac12 (n-m) - \sum_{x =-L/2}^{L/2} G_\downarrow(x) \cos k x, 
\end{align}
An $\exp(i k_0 x)/x^{p+1}$ term in $G_\downarrow(x)$ corresponds to a discontinuity
in the $p$'th derivative of $n_\downarrow(k)$ at $k=k_0.$ Thus $n_\downarrow(k)$ and 
its derivatives are discontinuous at $k=\pm k_{F\downarrow}$.
 In addition, the second and higher derivatives of 
$n_\downarrow(k)$ are discontinuous at 
$|k| = |k_{F\downarrow} \pm 2 k_{F\uparrow} |.$
The discontinuity at $k = k_{F \downarrow}$ is derived from Eq.(\ref{down-asymp}) as
\begin{align}
\Delta n_\downarrow (k_{F \downarrow}) \equiv n_\downarrow(k_{F \downarrow}+0)  -
 n_\downarrow(k_{F \downarrow}-0)   =   - \sqrt{(1-n)(1-m)} ,
\end{align}
while the up spin discontinuity is given in I as
\begin{align}
\Delta n_\uparrow (k_{F\uparrow}) = -\sqrt{(1-n)/(1-m)},
\end{align}
which agrees with  $\Delta n_\downarrow (k_{F \downarrow})$ with $m=0$.
Note that the majority spin ($\uparrow$) has a larger discontinuity,
giving $\Delta n_\uparrow (k_{F \uparrow})=1$
in the case of full polarization $m=n$.
This must be the case, since the system becomes the same as free fermions without spin.
The next singularity is the discontinuity in the slope, which is on top of $\Delta n_\downarrow (k_{F \downarrow})$ and is given by
\begin{align}
\Delta \left( \frac{d n_\downarrow (k_{F \downarrow })}{d k} \right)
 =   \frac{n_\downarrow}{2\pi (1-n)^{1/2}(1-m)^{1/2}} .
\label{dn/dk} 
\end{align}
Furthermore, the discontinuities in the curvature are obtained from Eq.(\ref{down-asymp}) as
\begin{eqnarray}
\Delta \left( \frac{d^2 n_\downarrow (2k_{F \uparrow }+k_{F \downarrow })}{d k^2} \right)
& = & +\frac{\sqrt{1-m}}{4\pi^2(1-n)^{3/2}} 
\label{3kF}\\
\Delta \left( \frac{d^2 n_\downarrow (2k_{F \uparrow }-k_{F \downarrow })}{d k^2} \right)
& = & -\frac{\sqrt{1-n}}{4\pi^2(1-m)^{3/2}}.
\label{2kF-kF}
\end{eqnarray}
In the case of singlet $m = 0$, Eq.(\ref{3kF}) corresponds to the momentum $k=3k_F$, while Eq.(\ref{2kF-kF}) contributes to the discontinuity in the curvature at $k=k_F$.
On the other hand, we have obtained in I:
\begin{equation}
\Delta \left( \frac{d ^2n_\uparrow (k_{F\uparrow}+2k_{F\downarrow})}{dk^2} \right) =
\frac{1}{4\pi^2(1-n)^{3/2}(1-m)^{1/2}}.
\label{curvature-3kF}
\end{equation}
Eqs.(\ref{curvature-3kF}) and (\ref{3kF}) are reduced to the same with $m=0$.

Note that there is no singularity in $n_\uparrow (k)$
at $k=\pm (k_{F\uparrow}-2k_{F\downarrow})$,
while there is one in $n_\downarrow (k)$
at $k=\pm (k_{F\downarrow}-2k_{F\uparrow})$.
This curious asymmetry between up and down spins can be understood by appealing to elementary excitations in the supersymmetric {\it t-J} model \cite{kuramoto91,ha,arikawa04b}. 
Namely, the fermionic excitations are combinations of charge excitations: holons and antiholons, and spin excitations: spinons and antispinons.  
The threshold momentum and the quantum number for each gapless excitation is given by \cite{ha,kuramoto-kato}
\begin{align}
&{\rm spinon: }\ \pm\pi m/2, \ {\rm with\ spin}\ 1/2,  &\
{\rm antispinon: }\ \pm\pi m, \ {\rm with\ spin}\ 1, \\
&{\rm holon: }\ \pm\pi n/2, \ {\rm with\ charge}\ +1, &\   
{\rm antiholon: }\ \pm\pi n, \ {\rm with\ charge}\  -2.
\end{align}
The simplest is a holon-spinon excitations which makes up a fermionic hole.  
The excitation gives the characteristic momenta,
$\pm k_{F\uparrow} = \pm \pi (n+m)/2$ and 
$\pm k_{F\downarrow} = \pm \pi (n-m)/2$.

On the other hand, an electron addition excitation has the charge $-1$, and must involve an antiholon.
The simplest combination consists of 
\begin{align}
{\rm (holon)+(spinon)+(antiholon)
}.
\end{align}
The threshold momenta are not only $\pm k_F$ but those which are reduced to $\pm 3k_F$ in the singlet limit.  For example, we obtain 
\begin{align}
\pi n/2+\pi m/2+\pi n = \pi n/2-\pi m/2+2\pi (n+m)/2 = k_{F\downarrow}+2k_{F\uparrow}.
\end{align}

Furthermore, 
the combination (holon)+(spinon)+(antispinon) 
can also make up a hole excitation with 
positive charge.  However the spin can take only the value $+1/2$, 
since the antispinon excitation from the magnon condensate accompanies the spin flip from down to up, but not the opposite.  
Because $n_\uparrow (k) = \langle c_{k\uparrow}^\dagger c_{k\uparrow}\rangle$ involves hole excitations with spin $-1/2$ as a result of acting $c_{k\uparrow}$ to the ground state, the antispinon cannot participate in the process.  In $n_\downarrow (k)$, on the other hand,
excitations accompany the spin change +1/2 by $c_{k\downarrow}$.  Then a threshold momentum is given by
\begin{align}
-\pi n/2-\pi m/2 -\pi m
= \pi (n-m)/2-2\pi (n+m)/2 = k_{F\downarrow}-2k_{F\uparrow}.
\end{align}
Hence there emerges a weak singularity at this momentum 
and the minus sign counterpart.

\begin{figure}
\includegraphics[width=0.59\linewidth,clip]{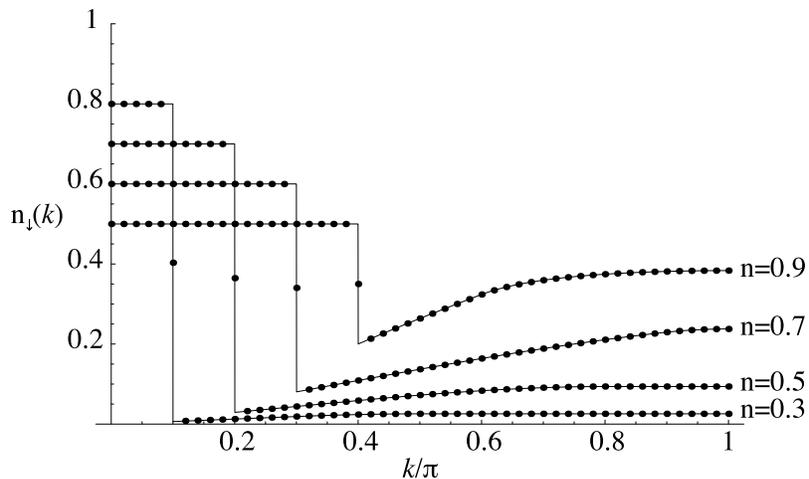}
\caption{
Plot of $n_\downarrow(k)$ versus $k$ 
with $n=0.9, 0.7, 0.5, 0.3$, and $m=0.1$ for all cases.
For comparison, the Kollar-Vollhardt results \cite{kollar} are also shown by solid lines.
}
\label{momentumdist}
\end{figure}

In order to see the global behavior in the momentum space,  we numerically carry out the Fourier transform of $G_\downarrow (x)$, and derive the momentum distribution function 
$n_\downarrow(k)$.   
The analytic calculation also seems possible, but is extremely complicated~\cite{arikawa07}.  
In this paper we are satisfied with numerical comparison with previous results \cite{kollar}.  As shown in Fig. \ref{momentumdist}, the agreement 
between the present and previous results
is excellent.
It is difficult to identify the tiny discontinuities in the curvature at
$k_{F\downarrow}-2k_{F\uparrow}$
in Fig.\ref{momentumdist}.
However, the discontinuity in the slope at $k_{F\downarrow}$ is clearly seen and is characterized by Eq.(\ref{dn/dk}).

\section{Concluding Remarks}

We have obtained the analytic form of the density matrix of the Gutzwiller
wave function with allowance of magnetization.   Together with the
previous results for the majority spin, we have now expressions for both components
of spins.  
The density matrix in the thermodynamic limit is characterized by
spin and charge velocities, which represent elementary
excitations in the supersymmetric {\it t-J} model with inverse-square
exchange and transfer \cite{kuramoto91,ha,kuramoto-kato}.  The singularities in the momentum distribution are identified completely.   The identification through the asymptotic form of the density matrix is more effective than previous methods \cite{ogata} which try to analyze them in the momentum space.

Although we have dealt with many complicated terms, the final result
seems relatively simple, as given by Eq.(\ref{main}). 
The simplification
emerges only in the thermodynamic limit. 
Similar simplification has also been seen in the
exact dynamics, where an extremely complicated expression for finite size
leads to a final result in terms of simple form factors \cite{arikawa04b}. 
In the present case, even after the thermodynamic limit is taken, there is considerable further
simplification, with double integrals involving products of four Bessel
functions eventually reducing to Eq.(\ref{main}).  
It is tempting to
speculate that there may be a way to work directly in the thermodynamic
limit without bothering about finite size, or even to obtain a completely
different real space formulation in which the final results are obtained
without `fortuitous' cancellations.

\acknowledgments
MA is supported by the University of Tsukuba Research Initiative.
Parts of the numerical calculations were carried out on the computers  
at YITP in Kyoto University.

%.  
\appendix
\section{Identities involving Bessel functions}

Here we summarize the properties of Bessel functions, which are used in this paper.
Bessel functions have the simple Wronskian:
\begin{align}
J_0 (x) Y_0'(x) - J_0' (x) Y_0(x) = 
\frac 2{\pi x}.
\label{Wronskian}
\end{align}
which comes from Abel's identity.  
Another important property is the differentials:
\begin{align}
Z_0'(x) = -Z_1(x), \quad
Z_1'(x) = Z_0(x)-\frac 1x Z_1(x),
\label{diff}
\end{align}
which apply for cases $Z_\nu(x)=J_\nu(x), \ Y_\nu (x)$, and their linear combinations with $\nu=0,1$.  
With Eq.(\ref{diff}) we obtain
\begin{equation}
\frac{d}{d x} \Bigg([Z_0(x) - i Z_1 (x)]x e^{i x}\Bigg) =
\frac{d}{d x} \Bigg([Z_0(x) + i Z_0' (x)]x e^{i x}\Bigg) =
Z_0(x) e^{i x} .
\end{equation}
Hence the right-hand side can be integrated.
We introduce the notation:
 $u(x)\equiv J_0(x)e^{ix}= U'(x)$ and $v(x) \equiv Y_0(x)e^{ix}=V'(x)$,
where
\begin{align}
U(x) = \left[ J_0(x)+iJ_0'(x) \right]xe^{ix}, \quad
V(x) = \left[ Y_0(x)+iY_0'(x) \right]xe^{ix}.
\end{align}
Then we consider the integral:
\begin{align}
K_{\alpha\beta}\equiv \int_a^b dx \ \alpha^\prime(x)\beta(x),
\end{align}
where $\alpha$ and $\beta$ denote either $U$ or $V$.
It is obvious that 
\begin{align}
K_{\alpha\alpha} = \frac 12 \left[ \alpha(b)^2 - \alpha(a)^2\right].
\end{align}
Therefore we obtain
\begin{align}
K_{uu} = \frac 12 \left[ U(b)^2 -U(a)^2\right], \quad
K_{vv} = \frac 12 \left[ V(b)^2 -V(a)^2\right].
\end{align}
The other components are manipulated as
\begin{align}
K_{uv} = \int_a^b dx U^\prime(x) V(x) = 
U(x)V(x)\big|_a^b -\int_a^b dx U(x) V'(x) = U(x)V(x)\big|_a^b - K_{vu},
\end{align}
which leads to 
\begin{align}
K_{uv}+K_{vu} = U(x)V(x)\big|_a^b.
\end{align}
On the other hand, the difference is organized as
\begin{align}
K_{uv}- K_{vu} 
&= \int_a^b dx J_0(x)e^{ix} xe^{ix}\left[ Y_0(x)+iY_0'(x) \right] -(J\leftrightarrow Y) \notag\\
&= i\int_a^b dx \left[ J_0(x)Y_0'(x) -Y_0(x)J_0'(x)  \right]xe^{2ix} \notag\\ 
&= \frac 1\pi \left( e^{2ib}-e^{2ia} \right)
,
\end{align}
where we have used the Wronskian Eq.(\ref{Wronskian}).
Hence we obtain
\begin{align}
K_{uv} 
&= \frac 1{2\pi} \left( e^{2ib}-e^{2ia} \right) +\frac 12
\left[ U(b) V(b) - U(a) V(a) \right] , \notag\\
K_{vu} 
&= \frac {-1}{2\pi} \left( e^{2ib}-e^{2ia} \right) +\frac 12
\left[ U(b) V(b) -U(a) V(a)\right]. \notag\\
\end{align}
By taking $a=v_cx/2$ and $b=v_sx/2$, the imaginary parts of $K_{\alpha\beta}$ gives the results for $I_{\alpha\beta}$ quoted in Eqs.(\ref{I_JJ}) -- (\ref{I_YJ}).

%.%%%%%%%%%%%%%%%%%%%%%%%%%%%%%%%%%%%%%%%


\begin{thebibliography}{99}
\bibitem{note1}
This condition was stated wrongly in I, and should be replaced by the present one.
\bibitem{note2}
The minus sign for the cosine term in this expression and in Eq.(78)  in I was incorrectly written as plus.
\end{thebibliography}
\end{document}